# Applications of local fractional calculus to engineering in fractal time-space:
# Local fractional differential equations with local fractional derivative

Yang XiaoJun

*Department of Mathematics and Mechanics, China University of Mining and Technology, Xuzhou Campus, Xuzhou, Jiangsu, 221008, P. R.C*

dyangxiaojun@163.com

This paper presents a better approach to model an engineering problem in fractal-time space based on local fractional calculus. Some examples are given to elucidate to establish governing equations with local fractional derivative.

## 1 Introduction

Local fractional calculus is a generalization of differentiation and integration of the functions defined on fractal sets. The idea of local fractional calculus has been a subject of interest not only among mathematicians but also among physicists and engineers [1-15].

There are many definitions of local fractional derivatives and local fractional integrals (also called fractal calculus) [1-13]. Hereby we write down Gao-Yang-Kang definitions as follows. Gao-Yang- Kang local fractional derivative is denoted by [7-8,10-11,14]

$$f^{(\alpha)}(x_0) = \frac{d^\alpha f(x)}{dx^\alpha}\bigg|_{x=x_0} = \lim_{x \to x_0} \frac{\Delta^\alpha (f(x) - f(x_0))}{(x - x_0)^\alpha}, \qquad (1.1)$$

with $\Delta^\alpha (f(x) - f(x_0)) \cong \Gamma(1+\alpha)\Delta(f(x) - f(x_0))$. Gao-Yang-Kang local fractional integral of $f(x)$ of order $\alpha$ in the interval $[a,b]$ is denoted by [7,9-11,14]

$$_aI_b^{(\alpha)} f(x) = \frac{1}{\Gamma(1+\alpha)} \int_a^b f(t)(dt)^\alpha = \frac{1}{\Gamma(1+\alpha)} \lim_{\Delta t \to 0} \sum_{j=0}^{j=N-1} f(t_j)(\Delta t_j)^\alpha, \qquad (1.2)$$

with $\Delta t_j = t_{j+1} - t_j$ and $\Delta t = \max\{\Delta t_1, \Delta t_2, \Delta t_j, \ldots\}$, where for $j = 0, \ldots, N-1$, $[t_j, t_{j+1}]$ is a partition of the interval $[a,b]$ and $t_0 = a, t_N = b$. Yang presented that there exists the relation [14]

$$|f(x) - f(x_0)| < \varepsilon^\alpha \tag{1.3}$$

with $|x - x_0| < \delta$, for $\varepsilon, \delta > 0$ and $\varepsilon, \delta \in \mathbb{R}$. Then $f(x)$ is called local fractional continuous on the interval $(a,b)$, denoted by

$$f(x) \in C_\alpha(a,b). \tag{1.4}$$

In this paper, our attempt is to model an engineering problem in fractal-time space based on local fractional calculus.

## 2 preliminaries

### 2.1 Notation

*Definition 1*

There always exists the relation

$$|f(x) - f(y)| < \zeta|x - y|^\alpha, \text{ for any } x, y \in \mathbb{R}, \tag{2.1}$$

where $\zeta$ is constant. $f(x)$ is called Hölder function of exponent of $\alpha$.

*Definition 2*

There always exists [14]

$$f(x) - f(x_0) = o\left((x - x_0)^\alpha\right) \text{ for any } x_0, x \in [a,b], \tag{2.2}$$

which is called $\alpha$ continuous, or local fractional continuous on the interval $[a,b]$.

**Remark 1.** From (2.1) and (2.2) this case is called unity local fractional continuity of $f(x)$. Form (1.3) this definition is called local fractional continuity of $f(x)$ (or called standard local fractional continuous), by denoted by

$$\lim_{x \to x_0} f(x) = f(x_0), \tag{2.3}$$

For $x \in (a,b)$, we have the notation

$$f(x) \in C_\alpha(a,b). \tag{2.4}$$

## 2.2 Generalized local fractional Taylor formula with local fractional derivative of one-variable function

*Proposition 1*

Suppose that $f(x) \in C_\alpha[a,b]$, then [4]

$$_aI_b^{(\alpha)}f(x) = f(\xi)\frac{(b-a)^\alpha}{\Gamma(\alpha+1)}, \quad a < \xi < b. \tag{2.5}$$

*Theorem 2*

Suppose that $f^{(k\alpha)}(x), f^{((k+1)\alpha)}(x) \in C_\alpha(a,b)$, for $0 < \alpha \leq 1$, then we have

$$_{x_0}I_x^{(k\alpha)}[f^{(k\alpha)}(x)] - {_{x_0}}I_x^{((k+1)\alpha)}[f^{((k+1)\alpha)}(x)] = f^{(k\alpha)}(\xi)\frac{(x-x_0)^{k\alpha}}{\Gamma(k\alpha+1)}, \tag{2.6}$$

with $a < x_0 < \xi < x < b$, where $_{x_0}I_x^{((k+1)\alpha)}f(x) = \overbrace{_{x_0}I_x^{(\alpha)}\ldots {_{x_0}}I_x^{(\alpha)}}^{k+1 \text{ times}} f(x)$ and

$$f^{((k+1)\alpha)}(x) = \overbrace{D_x^{(\alpha)}\ldots D_x^{(\alpha)}}^{k+1 \text{ times}} f(x).$$

Proof. From (2.5) we have

$$_{x_0}I_x^{((k+1)\alpha)}[f^{((n+1)\alpha)}(x)] = {_{x_0}}I_x^{(k\alpha)}\left[\frac{1}{\Gamma(1+\alpha)}\int_{x_0}^x f^{((n+1)\alpha)}(x)(dt)^\alpha\right] \tag{2.7}$$

$$= {_{x_0}}I_x^{(k\alpha)}\left(f^{(k\alpha)}(x) - f^{(k\alpha)}(\xi)\right) \tag{2.8}$$

$$= {_{x_0}}I_x^{(k\alpha)}f^{(k\alpha)}(x) - {_{x_0}}I_x^{(k\alpha)}f^{(k\alpha)}(\xi). \tag{2.9}$$

Successively, it follows from (2.9) that

$$_{x_0}I_x^{(k\alpha)}f^{(k\alpha)}(\xi) = f^{(k\alpha)}(\xi) \, {_{x_0}}I_x^{(k\alpha)}1 \tag{2.10}$$

$$= f^{(k\alpha)}(\xi) \, {_{x_0}}I_x^{((k-1)\alpha)}\left[\frac{1}{\Gamma(1+\alpha)}(x-x_0)^\alpha\right]$$

$$(2.11)$$

$$= f^{(k\alpha)}(\xi) \, {_{x_0}}I_x^{((k-2)\alpha)}\left[\frac{\Gamma(1+\alpha)}{\Gamma(1+2\alpha)} \bullet \frac{1}{\Gamma(1+\alpha)}(x-x_0)^{2\alpha}\right]$$

$$(2.12)$$

$$= f^{(k\alpha)}(\xi)\frac{(x-x_0)^{k\alpha}}{\Gamma(k\alpha+1)}$$

$$(2.13)$$

Hence we have the result.

*Theorem 3 (Generalized mean value theorem for local fractional integrals)*

Suppose that $f(x) \in C_\alpha [a,b]$, $f^{(\alpha)}(x) \in C(a,b)$, we have

$$f(x) - f(x_0) = f^{(\alpha)}(\xi) \frac{(x-x_0)^\alpha}{\Gamma(\alpha+1)}, \quad a < x_0 < \xi < x < b. \tag{2.14}$$

*Proof.* Taking $k = 1$ in (2.6), we deduce the result.

*Theorem 4 (Generalized local fractional Taylor formula)*

Suppose that $f^{((k+1)\alpha)}(x) \in C_\alpha(a,b)$, for $k = 0,1,...,n$ and $0 < \alpha \le 1$, then we have

$$f(x) = \sum_{k=0}^{n} \frac{f^{(k\alpha)}(x_0)}{\Gamma(1+k\alpha)} (x-x_0)^{k\alpha} + \frac{f^{((n+1)\alpha)}(\xi)}{\Gamma(1+(n+1)\alpha)} (x-x_0)^{(n+1)\alpha} \tag{2.15}$$

with $a < x_0 < \xi < x < b$, $\forall x \in (a,b)$, where $f^{((k+1)\alpha)}(x) = \overbrace{D_x^{(\alpha)}...D_x^{(\alpha)}}^{k+1 \ times} f(x)$.

*Proof.* Form (2.6), we get

$$_{x_0}I_x^{(k\alpha)}[f^{(k\alpha)}(x)] - {_{x_0}I_x^{((k+1)\alpha)}}[f^{((k+1)\alpha)}(x)] = f^{(k\alpha)}(a) \frac{(x-x_0)^{k\alpha}}{\Gamma(k\alpha+1)}. \tag{2.16}$$

Successively, it follows from (2.16) that

$$\sum_{k=0}^{n} \left( {_{x_0}I_x^{(k\alpha)}}[f^{(k\alpha)}(x)] - {_{x_0}I_x^{((k+1)\alpha)}}[f^{((k+1)\alpha)}(x)] \right) = f(x) - {_{x_0}I_x^{((n+1)\alpha)}}[f^{((n+1)\alpha)}(x)] \tag{2.17}$$

$$= \sum_{k=0}^{n} f^{(k\alpha)}(x_0) \frac{(x-x_0)^{k\alpha}}{\Gamma(k\alpha+1)}. \tag{2.18}$$

By using (2.5) and (2.18) we have

$$_{x_0}I_x^{((n+1)\alpha)}\left[ f^{((n+1)\alpha)}(x) \right] = \frac{1}{\Gamma(1+\alpha)} \int_{x_0}^{x} {_aI_{x_0}^{(n\alpha)}} f^{((n+1)\alpha)}(x) (dt)^\alpha \tag{2.19}$$

$$= \frac{{_aI_{x_0}^{(n\alpha)}}\left[ f^{((n+1)\alpha)}(\xi)(x-x_0)^\alpha \right]}{\Gamma(1+\alpha)} \tag{2.20}$$

$$= f^{((n+1)\alpha)}(\xi) \frac{{_aI_{x_0}^{(n\alpha)}}(x-x_0)^\alpha}{\Gamma(1+\alpha)} \tag{2.21}$$

$$= \frac{f^{((n+1)\alpha)}(\xi)(x-x_0)^{(n+1)\alpha}}{\Gamma(1+(n+1)\alpha)} \tag{2.22}$$

with $a < x_0 < \xi < x < b$, $\forall x \in (a,b)$.

Combing the formulas (2.22) and (2.18) in (2.16), we have the result.

Hence, the proof of the theorem is completed.

## 2.3 Local fractional continuity of two-variable function

*Definition 3*

There exists the relation

$$|f(x,y) - f(x_0, y_0)| < \varepsilon^\alpha \tag{2.23}$$

with $|x - x_0| < \delta_1$ and $|y - y_0| < \delta_2$, for $\varepsilon, \delta_1, \delta_2 > 0$ and $\varepsilon, \delta_1, \delta_2 \in \mathbb{R}$. Then $f(x,y)$ is called local fractional continuous at $(x_0, y_0) \in E$. The function $f(x,y)$ is local fractional continuous in the region $E$, denoted by

$$f(x) \in C_\alpha(E). \tag{2.24}$$

*Definition 4*

Setting $f(x,y) \in C_\alpha(E)$, local fractional partial derivative of $f(x,y)$ with respect to $x$, is denoted by

$$\left.\frac{\partial^\alpha f(x,y)}{\partial x^\alpha}\right|_{x=x_0} = \lim_{x \to x_0} \frac{\Delta^\alpha(f(x,y) - f(x_0,y))}{(x-x_0)^\alpha} \tag{2.25}$$

with $\Delta^\alpha(f(x,y) - f(x_0,y)) \cong \Gamma(1+\alpha)\Delta(f(x,y) - f(x_0,y))$.

Similarly, local fractional partial derivative of $f(x,y)$ with respect to $y$, is denoted by

$$\left.\frac{\partial^\alpha f(x,y)}{\partial y^\alpha}\right|_{y=y_0} = \lim_{y \to y_0} \frac{\Delta^\alpha(f(x,y) - f(x,y_0))}{(y-y_0)^\alpha}, \tag{2.26}$$

with $\Delta^\alpha(f(x,y) - f(x,y_0)) \cong \Gamma(1+\alpha)\Delta(f(x,y) - f(x,y_0))$.

The second derivatives are denoted by

$$\frac{\partial^\alpha}{\partial x^\alpha}\left(\frac{\partial^\alpha f(x,y)}{\partial x^\alpha}\right) = \frac{\partial^{2\alpha} f(x,y)}{\partial x^\alpha \partial x^\alpha} = f_{xx}^{2\alpha}, \tag{2.27}$$

$$\frac{\partial^\alpha}{\partial y^\alpha}\left(\frac{\partial^\alpha f(x,y)}{\partial y^\alpha}\right) = \frac{\partial^{2\alpha} f(x,y)}{\partial y^\alpha \partial y^\alpha} = f_{yy}^{2\alpha}, \tag{2.28}$$

$$\frac{\partial^\alpha}{\partial x^\alpha}\left(\frac{\partial^\alpha f(x,y)}{\partial y^\alpha}\right) = \frac{\partial^{2\alpha} f(x,y)}{\partial x^\alpha \partial y^\alpha} = f_{yx}^{2\alpha}, \quad (2.29)$$

$$\frac{\partial^\alpha}{\partial y^\alpha}\left(\frac{\partial^\alpha f(x,y)}{\partial x^\alpha}\right) = \frac{\partial^{2\alpha} f(x,y)}{\partial y^\alpha \partial x^\alpha} = f_{xy}^{2\alpha}. \quad (2.30)$$

**Remark 2.** Suppose that $f_{yx}^{2\alpha}$ and $f_{xy}^{2\alpha}$ are local fractional continuous in a fractal region $R$ of the $xy$-plane, then

$$f_{yx}^{2\alpha} = f_{xy}^{2\alpha}. \quad (2.31)$$

For local fractional partial derivative of high order, we get

$$\frac{\partial^{n\alpha}}{\partial x^{n\alpha}} f = \overbrace{\frac{\partial^\alpha}{\partial y^\alpha} \frac{\partial^\alpha}{\partial y^\alpha} \bullet \ldots \bullet \frac{\partial^\alpha}{\partial y^\alpha}}^{n\alpha} f \quad (2.32)$$

and

$$\frac{\partial^{m+n\alpha} f}{\partial x^{n\alpha} \partial y^{m\alpha}} = \left(\overbrace{\frac{\partial^\alpha}{\partial x^\alpha} \frac{\partial^\alpha}{\partial x^\alpha} \bullet \ldots \bullet \frac{\partial^\alpha}{\partial x^\alpha}}^{n\alpha} \overbrace{\frac{\partial^\alpha}{\partial y^\alpha} \frac{\partial^\alpha}{\partial y^\alpha} \bullet \ldots \bullet \frac{\partial^\alpha}{\partial y^\alpha}}^{m\alpha}\right) f. \quad (2.33)$$

In similar manner, we get generalized local fractional Taylor series for two-variable function as follows:

Suppose that $f(x,y) \in C_\alpha(E)$, $\frac{\partial^{n\alpha}}{\partial x^{n\alpha} x^{(n-i)\alpha}} f \in C_\alpha(E)$, then

$$f(x,y) = \sum_{n=0}^{n=\infty} \sum_{i=0}^{i=n} (x-x_0)^{i\alpha} (y-y_0)^{(n-i)\alpha} \frac{\partial^{n\alpha}}{\partial x^{i\alpha} y^{(n-i)\alpha}} f(x_0, y_0), \quad (2.34)$$

with $(x_0, y_0) \in E$.

## 2.4 Applications of approximation of functions

From (2.34) we have the following relation

$$f(x,y) = \sum_{n=0}^{n} \sum_{i=0}^{i=n} (x-x_0)^{i\alpha} (y-y_0)^{(n-i)\alpha} \frac{\partial^{n\alpha}}{\partial x^{i\alpha} y^{(n-i)\alpha}} f(x_0, y_0) + R_{n\alpha}$$

$$(2.35)$$

where its reminder is

$$R_{n\alpha} = \sum_{n}^{n=\infty} \sum_{i=0}^{i=\infty} (x-x_0)^{i\alpha} (y-y_0)^{(n-i)\alpha} \frac{\partial^{n\alpha}}{\partial x^{i\alpha} y^{(n-i)\alpha}} f(x_0, y_0) \quad (2.36)$$

and

$$\lim_{n \to \infty} R_{n\alpha} = 0. \tag{2.37}$$

Now taking the formula (2.15) into account we arrive at this relation

$$E_\alpha(x^\alpha) = \sum_{k=0}^{\infty} \frac{x^{\alpha k}}{\Gamma(1+k\alpha)}. \tag{2.38}$$

which yields

$$E_\alpha(x^\alpha) \cong \sum_{k=0}^{n} \frac{x^{\alpha k}}{\Gamma(1+k\alpha)} \tag{2.39}$$

In similar manner, from (2.34) we get

$$E_\alpha((x+y)^\alpha) = \sum_{k=0}^{\infty} \frac{x^{\alpha k}}{\Gamma(1+k\alpha)} \sum_{k=0}^{\infty} \frac{x^{\alpha k}}{\Gamma(1+k\alpha)}. \tag{2.40}$$

which deduces

$$E_\alpha((x+y)^\alpha) = \sum_{k=0}^{n} \frac{x^{\alpha k}}{\Gamma(1+k\alpha)} \sum_{k=0}^{n} \frac{x^{\alpha k}}{\Gamma(1+k\alpha)}. \tag{2.41}$$

## 2.5 Useful results

Here the following formulas hold:

$$\frac{d^\alpha x^{k\alpha}}{dx^\alpha} = \frac{\Gamma(1+k\alpha)}{\Gamma(1+(k-1)\alpha)} x^{(k-1)\alpha}, \tag{2.42}$$

$$\frac{d^\alpha E_\alpha(kx^\alpha)}{dx^\alpha} = k E_\alpha(kx^\alpha); \tag{2.43}$$

$$\frac{d^\alpha E_\alpha(x^\alpha)}{dx^\alpha} = E_\alpha(x^\alpha). \tag{2.44}$$

# 3 Application to governing equation in engineering in fractal space

In this section some models for governing equations in fractal space are suggested. We start with typical models with local fraction derivative in engineering.

## 3.1 Some typical models with local fractional derivatives in engineering

*Model 1.* Local fractional transient heat conduction equation

The transient heat conduction equation in fractal spaces can be described by the equation

$$\frac{\partial^{2\alpha} y(x,t)}{\partial x^{2\alpha}} = \alpha \frac{\partial^{\alpha} y(t)}{\partial t^{\alpha}}, \quad t > 0, -\infty < x < +\infty, 0 < \alpha < 1, \tag{3.1}$$

with the initial and boundary conditions

$$\begin{cases} \dfrac{\partial^{\alpha} y(x,t)}{\partial x^{\alpha}} = 0, x = 0; \\ k\dfrac{\partial^{\alpha} y(x,t)}{\partial x^{\alpha}} + h\left(y\{x,t\} - y_{\infty}\right) = 0, x = L; \\ y(x,t) = y_i, t = 0 \end{cases} \tag{3.2}$$

where $\alpha$ is the fractal thermal diffusivity and $k$ is the thermal conductivity of the fractal wall material.

*Model 2.* Local fractional wave equation

The wave motion in fractional space can be described by the equation

$$\frac{\partial^{\alpha} y(t,x)}{\partial t^{\alpha}} = \frac{\partial^{2\alpha} y(t,x)}{\partial x^{2\alpha}}, 0 < \alpha < 1, \tag{3.3}$$

with the initial and boundary conditions

$$\begin{cases} \lim_{x \to +\infty} y(t,x) = 0 \\ \lim_{x \to -\infty} y(t,x) = 0 \\ \dfrac{\partial^{\alpha} y(t,x)}{\partial t^{\alpha}}\bigg|_{t=0} = \varphi(x). \end{cases} \tag{3.4}$$

*Model 3.* Local fractional diffusion equation

The diffusion equation in fractional space can be described by the equation

$$\frac{\partial^{\alpha} y(t,x)}{\partial t^{\alpha}} = a^{2\alpha} \frac{\partial^{2\alpha} y(t,x)}{\partial x^{2\alpha}}, t > 0, -\infty < x < +\infty, 0 < \alpha < 1, \tag{3.5}$$

with the boundary-value problem

$$\begin{cases} \lim_{x \to +\infty} y(t,x) = 0 \\ \lim_{x \to -\infty} y(t,x) = 0 \\ \dfrac{\partial^{\alpha} y(t,x)}{\partial t^{\alpha}}\bigg|_{t=0} = \phi(x). \end{cases} \tag{3.6}$$

### 3.2 Application to local fractional relaxation equation

The relaxation equation in fractional space can be described by the equation

$$\frac{\partial^{\alpha} y(t)}{\partial t^{\alpha}} + c^{\alpha} y(t) = 0, \quad c > 0, t > 0, 1 > \alpha > 0, \tag{3.7}$$

with the initial value $y(t)\big|_{t=0} = y_0$.

Its solution can be easily obtained, which reads

$$y(t) = y_0 E_\alpha\left(-c^\alpha t^\alpha\right). \tag{3.8}$$

From (2.44), taking $k = -c^\alpha$ implies that

$$\frac{d^\alpha E_\alpha\left(-c^\alpha t^\alpha\right)}{dx^\alpha} = -c^\alpha E_\alpha\left(-c^\alpha t^\alpha\right) \tag{3.9}$$

It follows from (3.9) that a special solution to equation (3.7)

$$y(t) = E_\alpha\left(-c^\alpha t^\alpha\right). \tag{3.10}$$

Hence we have the following relation

$$y(t) = m E_\alpha\left(-c^\alpha t^\alpha\right). \tag{3.11}$$

Taking initial value condition into account in (3.11), we obtain the solution of equation (3.7), which is

$$y(t) = y_0 E_\alpha\left(-c^\alpha t^\alpha\right). \tag{3.12}$$

The solution depends upon fractal dimension $\alpha$.

Given any point $t = t_0$, we have the local fractional Taylor-Yang expansion of $y(t)$ in the form

$$E_\alpha\left(-c^\alpha t^\alpha\right) = \sum_{k=0}^{\infty} (-1)^k \frac{c^{k\alpha} E_\alpha\left(-c^\alpha t_0^\alpha\right)(t-t_0)^{\alpha k}}{\Gamma(1+k\alpha)} \tag{3.13}$$

and we always get the Hölder relation

$$\left|E_\alpha\left(-c^\alpha t_1^\alpha\right) - E_\alpha\left(-c^\alpha t_2^\alpha\right)\right| \leq m\left|t_1 - t_2\right|^\alpha, \tag{3.14}$$

for any $t_1, t_2 > 0$.

Hence, from (3.13) and (3.14) its solution is local fractional continuous and exists fractal property.

## 4 Conclusions

The suggested governing equations with local fractional derivative are easy to be used for any fractal process and we get the solution to relaxation equation in fractional space. However, solutions to transient heat conduction equation, wave equation and diffusion equation in fractal space are much needed. Maybe, Yang-Laplace transforms [14] in fractal space get the result.